\documentstyle[12pt,ssi]{article}
\begin{document}
\input epsf

\begin{tabular}{lr}
Visual Techniques Laboratory & \hspace{1.25truein} UWSEA PUB 94-07\\
Department of Physics, FM-15 &\\
University of Washington &\\
Seattle, WA 98195 USA &\\
\end{tabular}

\title{DUMAND and AMANDA: High Energy Neutrino Astrophysics
\thanks{To be published in {\it Proc. SLAC Summer Institute 1994.}} }

\author{R. Jeffrey Wilkes\thanks{Supported by DOE Grant DE-FG06-91ER40614.}\\
Department of Physics, FM-15 \\
University of Washington, Seattle, WA 98195 \\[0.4cm]
}

\maketitle

\begin{abstract}

The field of high energy neutrino astrophysics is entering an
exciting new phase as two new large-scale observatories prepare to
come on line. Both DUMAND (Deep Underwater Muon and Neutrino
Detector) and AMANDA (Antarctic Muon and Neutrino Detector) had
major deployment efforts in 12/93--1/94. Results were mixed, with
both projects making substantial progress, but encountering
setbacks that delayed full-scale operation. The  achievements,
status, and plans (as of 10/94) of these two projects will be
discussed.

\end{abstract}

\section{Introduction}

Nature must love neutrinos, because she makes so many of them:
neutrinos are more abundant than photons (about $10^3$/cm$^3$;
10$^{17}$/sec pass through your body). In addition to the enormous
density of big-bang relict neutrinos, effectively undetectable due
to their tiny energy, neutrinos are produced copiously at solar
(few MeV) and astrophysical (GeV--EeV) energy scales by a variety
of processes.

Since neutrinos are uncharged, (probably) massless leptons, they
interact with matter only via the weak force. Thus, while they
share some features with photons as a probe of the distant universe
(straight-line propagation from sources at the speed of light),
they offer the advantage of being able to penetrate regions with
moderate mass density such as the center of our Galaxy. Neutrinos
therefore let us observe regions of the universe as yet unseen.
High energy photons and neutrinos are produced by similar
processes, for example by the decay of mesons produced in hadronic
interactions of charged particles near a cosmic ray source. Compact
binary systems, in which a neutron star orbits a giant companion,
are excellent candidates for copious photon and neutrino
production, as protons are accelerated in the pulsar's intense,
rapidly changing magnetic fields, and interact in the periphery of
the companion star. One therefore expects to see neutrinos from
sources that produce high energy gamma rays. While current
experiments have seen clear gamma ray signals from only a few
identifiable point sources\cite{haines 1994} this almost
certainly must be due to experimental limitations. We know that cosmic
ray hadrons (protons and/or nuclei) are produced in the EeV
(10$^{18}$ eV) region, beyond the reasonable limit for supernova shock
acceleration (thought to account for most of the cosmic ray flux
below 10$^{15}$ eV) and if protons are accelerated there must be
interactions near the sources yielding photons and neutrinos.

\begin{figure}
\epsfxsize=5.75truein
\epsffile{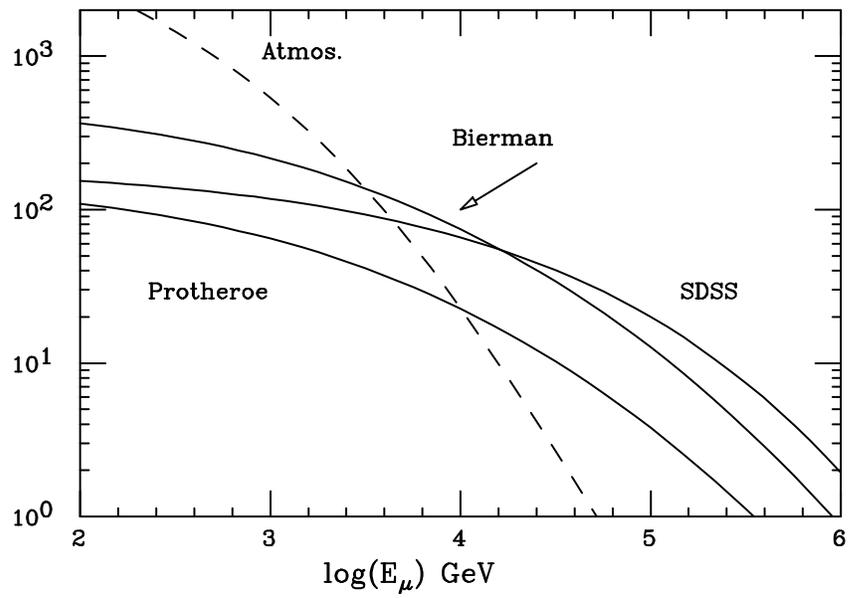}
\caption{Expected rates (events per year) in DUMAND-II from AGN neutrinos from
several leading models.}
\label{agnmuvse}
\end{figure}

Other source mechanisms are unique to neutrinos, such as the widely
accepted models for abundant UHE neutrino production in Active
Galactic Nuclei (AGNs)\cite{stecker 1991}. Here the power source is
thought to be a black hole about 10$^{6-9}$ times as massive as the
Sun, protons are accelerated by shocks in jets or flow in the
accretion disk, and neutrinos are produced by interactions with the
high density of UV or optical photons near the nucleus. Model
calculations show that we should expect a neutrino spectrum much
harder than the normal cosmic ray spectrum, leading to a
previously-unexpected wealth of neutrinos in the PeV (10$^{15}$ eV)
range (Fig.~\ref{agnmuvse}), making practical secondary studies such as
tomography of the
Earth's core. Neutrino observations in the GeV--PeV range thus complement
photon
observations at all energies, and provide useful
discrimination between some models\cite{HENA review}\cite{stenger
1992}.

There are basic physics questions to be answered: why do neutrinos
come in three flavors, do they have mass, are they the solution to
the dark-matter puzzle. As an example, recent results from the
Kamiokande-III\cite{kamiokande 1994} and IMB\cite{IMB 1993}
underground neutrino detectors suggest a substantial deviation from
expectation in the observed ratio of muon to electron neutrinos
produced in the atmosphere; it is possible to interpret the data in
terms of neutrino oscillations, consistent with an island of
allowed values in the mixing-angle/mass-difference parameter space.
Neutrino astrophysics experiments like these provide a way to
address such questions with costs an order of magnitude below those
of contemporary accelerator experiments ({\it ie,} on the order of
US\$ 10 million). There
is no question
that in future we will have to find ways to do particle physics that
make much smaller demands on the world economy.

But for many of us, one of the most attractive features of neutrino
astrophysics is the virginity of the field: the unexpected is
always a possibility, and historically science has made great
advances whenever a new mode of viewing the universe has been
tested. Perhaps the first large-scale neutrino detectors will
eventually have the significance of Galileo's spyglass.

\begin{figure}
\epsfxsize=5.75truein
\epsffile{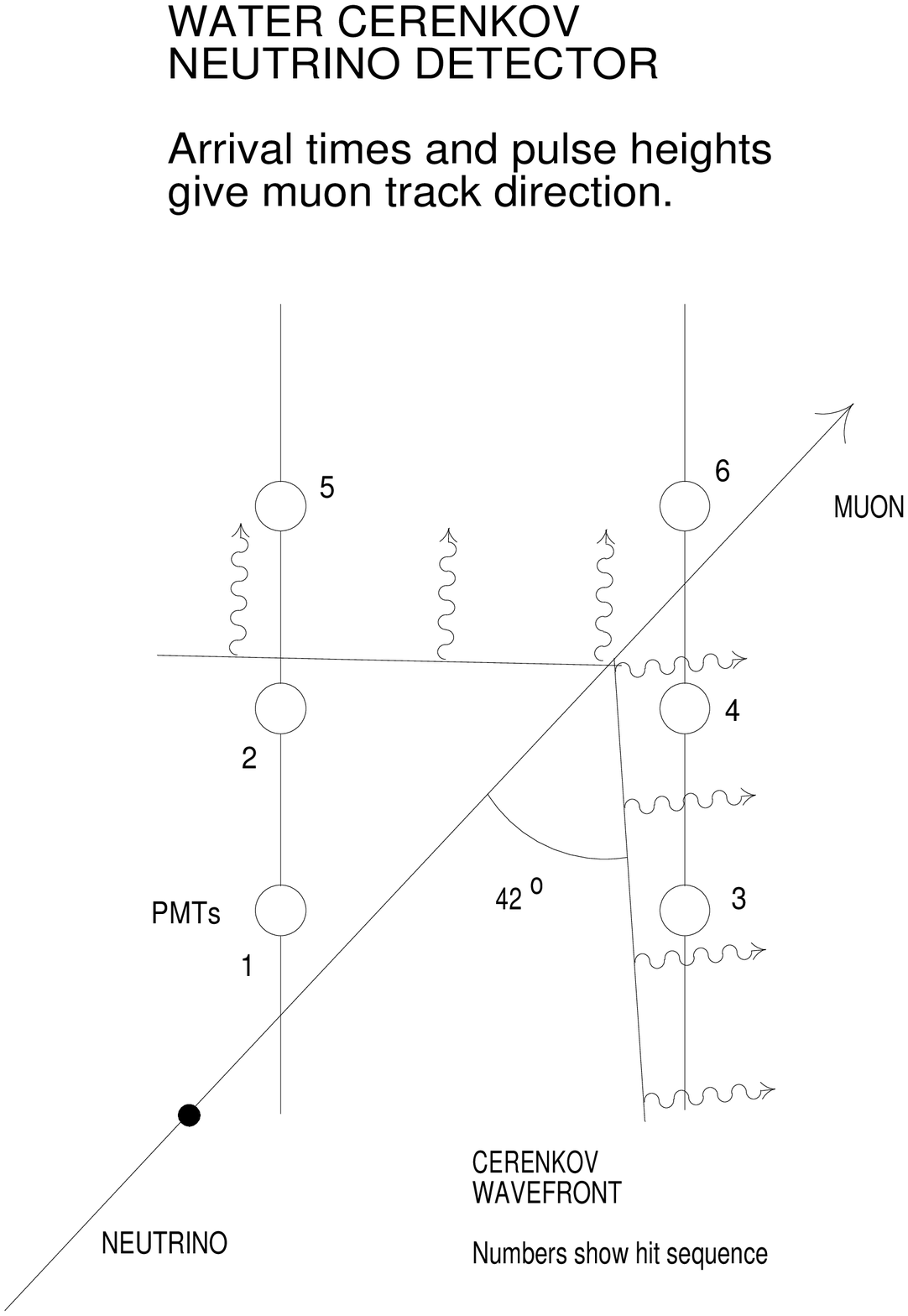}
\caption{Water (or ice) \v{C}erenkov detector concept.}
\label{watercer}
\end{figure}

\begin{figure}
\epsfxsize=5.75truein
\epsffile{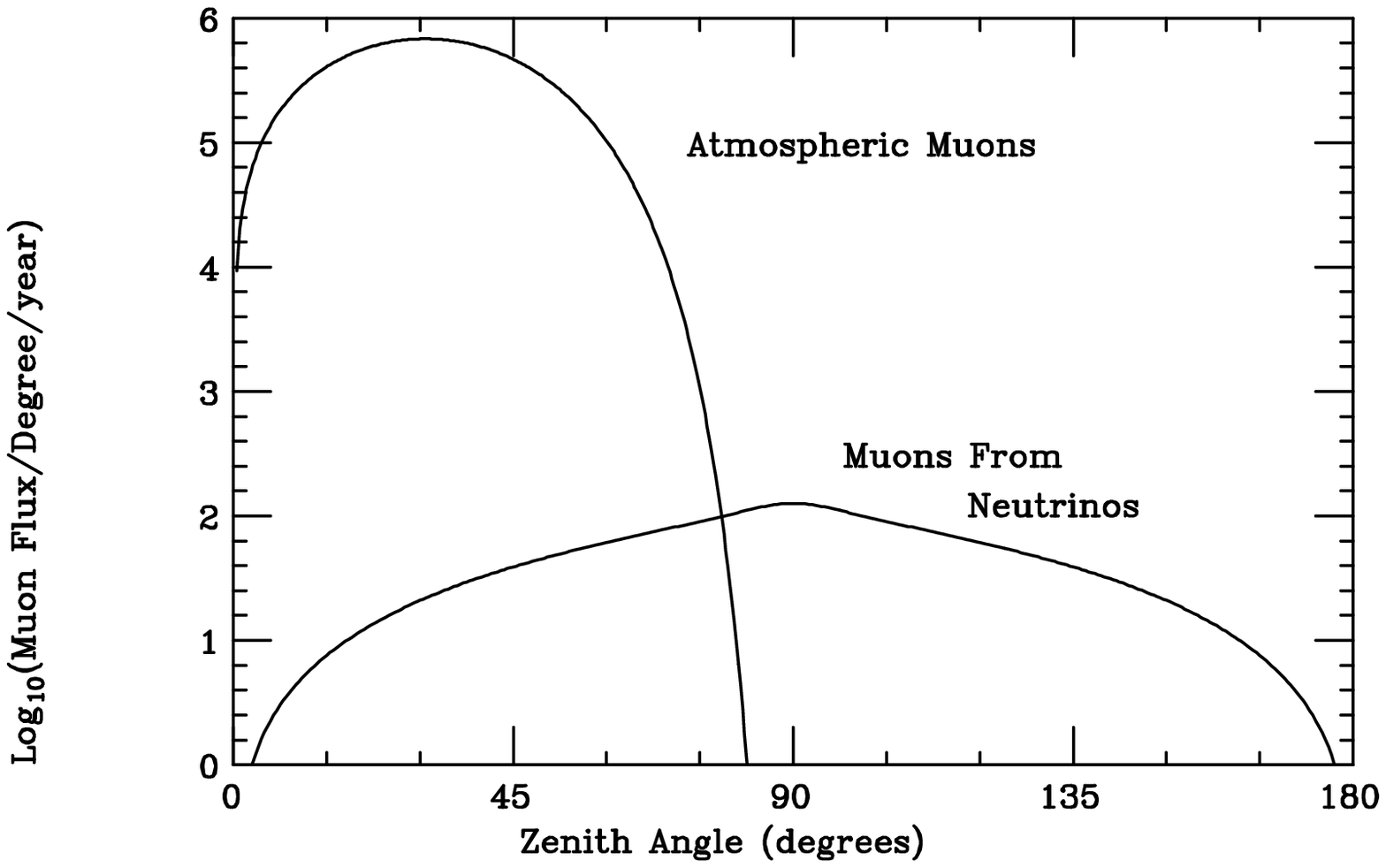}
\caption{Muon angular distribution: background muons from atmospheric cosmic
ray
interactions are cut off by looking only at upward-moving tracks.}
\label{muangdist}
\end{figure}

The basic concept of a water or ice \v{C}erenkov detector is
illustrated in Fig.~\ref{watercer}, which depicts a neutrino interaction
producing a muon. Seawater (or ice) serves a triple purpose, acting
as a low-cost massive target, supplying a track sensitive,
transparent medium for production and propagation of Cerenkov
radiation by charged particles, and also providing a thick, uniform
overburden (in contrast to underground experiments, with nonuniform
material and an irregular surface profile) to filter out downward-moving
background particles. The water volume is instrumented with
an array of sensitive photomultiplier tubes (PMTs). The attenuation
length for light in water at the DUMAND site in the appropriate
wavelength range is about 40 m, which defines the scale of the
transverse spacing of detector ``strings", and the vertical
separation of PMTs is set at 10m to provide adequate photocathode
coverage; similar parameters apply to ice. Upward moving neutrinos,
having passed through the earth (and thus being accompanied by
essentially no background, as shown in Fig.~\ref{muangdist}), interact in the
contained volume of water or in the nearby seabed, producing muons, charged
particles moving near the speed of light {\it in vacuo}, which will therefore
generate Cerenkov radiation in the water (n=1.35 in seawater). The
Cerenkov light is produced in a characteristic cone-shaped pattern,
and thus information on the arrival time and pulse intensity
recorded at each of the photomultiplier tubes can be used to
reconstruct the muon track direction. For energetic muons,
collected photoelectron statistics can be sufficient to provide a
muon energy estimate. In the case of ``contained events", where the
event vertex is within the sensitive volume, the hadron-electromagnetic
cascade can be observed and a more accurate energy
estimate made.

The idea of detecting high energy astrophysical neutrinos is an old
one, and calls for development of a practical detector date from at
least the early 60s\cite{markov 1960}. Apparent anomalies in the
underground muon flux\cite{keuffel 1971} stimulated interest in
underwater muon detectors offering uniform overburden, and
indirectly fostered development of the current generation of large-scale
neutrino detectors\cite{learned 1967}. The DUMAND concept
in more or less its present form has been discussed, and
construction projects of various degrees of practicality have been
proposed, since the mid-70s\cite{DUMAND 1976}. The water Cerenkov
technique was further refined in the early 80s by the successful
construction and operation of large-scale proton-decay detectors
(later used as low-energy neutrino observatories) by the
IMB\cite{IMB 1992} and Kamiokande\cite{kamiokande 1992} Collaborations. These
projects made it possible for the DUMAND proposal to be accepted
for construction funding by the US Department of Energy in
1990. The cost and risks involved in deep-ocean engineering
operations were still a matter of concern. At about the same time
the AMANDA group proposed an alternative approach, in which the
Antarctic ice cap replaces the ocean as overburden, target and
detecting medium. Deployment operations take place from the stable
platform of the South Pole Station. AMANDA has its substantial
logistical requirements covered by the US National Science
Foundation's Office of Polar Programs, which supports all
scientific research operations in Antarctica.

The remainder of this article will compare and contrast AMANDA and
DUMAND, ending with a look at initiatives presently being
undertaken for the next step in detector sensitivity, a second-generation
observatory of scale 1 km$^3$.  As a participant in DUMAND,
I hope to avoid any inadvertent
bias in this review.
Two parallel efforts in Europe, the NESTOR project in Greece and
the Baikal project in Russia, will not be discussed here simply due
to lack of space. Both projects are making significant progress and
will have important effects on the development of this rapidly-growing field.

\section{DUMAND}

\begin{figure}
\epsfxsize=5.75truein
\epsffile{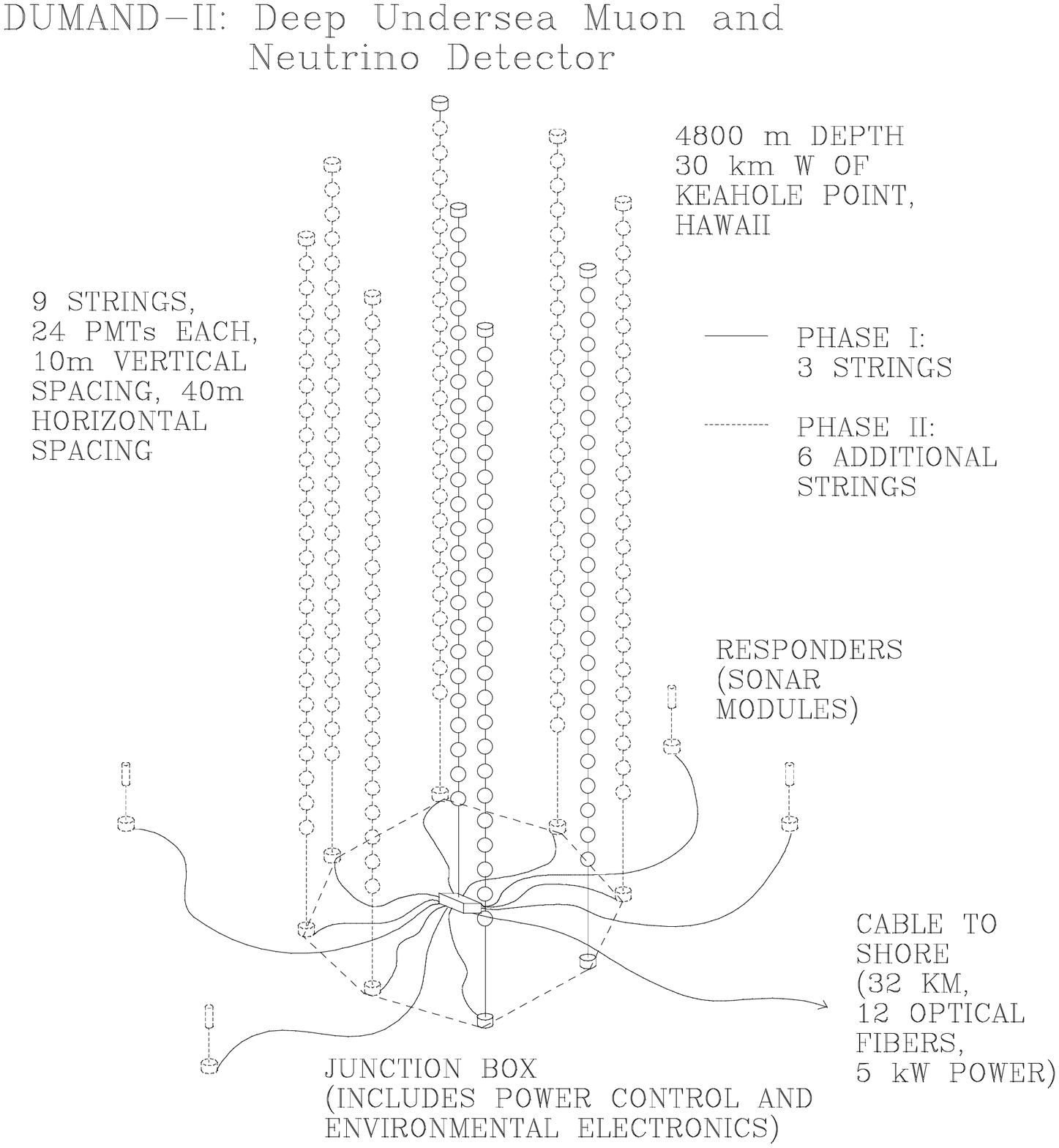}
\caption{DUMAND-II underwater neutrino detector array.}
\label{dumarray9}
\end{figure}

Taking our subjects in order of age, the DUMAND project has been
discussed in one form or another for nearly 30 years\cite{roberts 1992}.
The detector presently being constructed in Hawaii is called
DUMAND-II\cite{dumandcollab}. DUMAND-I refers to a ship-suspended
single prototype string which was successfully operated in
1987\cite{babson 1990}. The funding plan provides for deployment of the
full 9-string array (Fig.~\ref{dumarray9}) in two phases: first 3 strings (the
triad) as a demonstration, and the remaining 6 strings (complete
octagon, plus center string) after about 1 year of testing and
operation. Details of the detector design and physics capabilities
have been published elsewhere\cite{icrcdumand 1993}.

The Island of Hawaii was selected for a variety of compelling
reasons: exceptional water clarity, proximity of an abyssal plain
with appropriate seabed characteristics to a suitable shore site
(30 km away), presence of an active particle physics group at the
nearby University of Hawaii in Honolulu, and pre-existing
laboratory infrastructure at the shore site, due to an ocean
thermal energy research project. The latter feature even provided
a cost-free conduit for the DUMAND shore cable to pass beneath the
surf zone, since the thermal energy project involves slant drilling
of tunnels into the ocean.

When completed, DUMAND-II will be an array of 216 Optical Modules
(OMs: photomultiplier tubes plus front-end electronics, encased in
a standard glass oceanographic pressure sphere) deployed on nine
vertical strings, which are moored in an octagonal pattern with 40m
sides and one string in the center (Fig.~\ref{dumarray9}). The instrumented
portion of each string begins 100m above the ocean floor to avoid
boundary-layer effects. In addition to OMs, the strings include
sets of hydrophones and other acoustical equipment, and calibration
modules, in which a constant output laser light source is used to
excite a scintillator ball viewed by the PMTs.

\begin{figure}
\epsfxsize=5.75truein
\epsffile{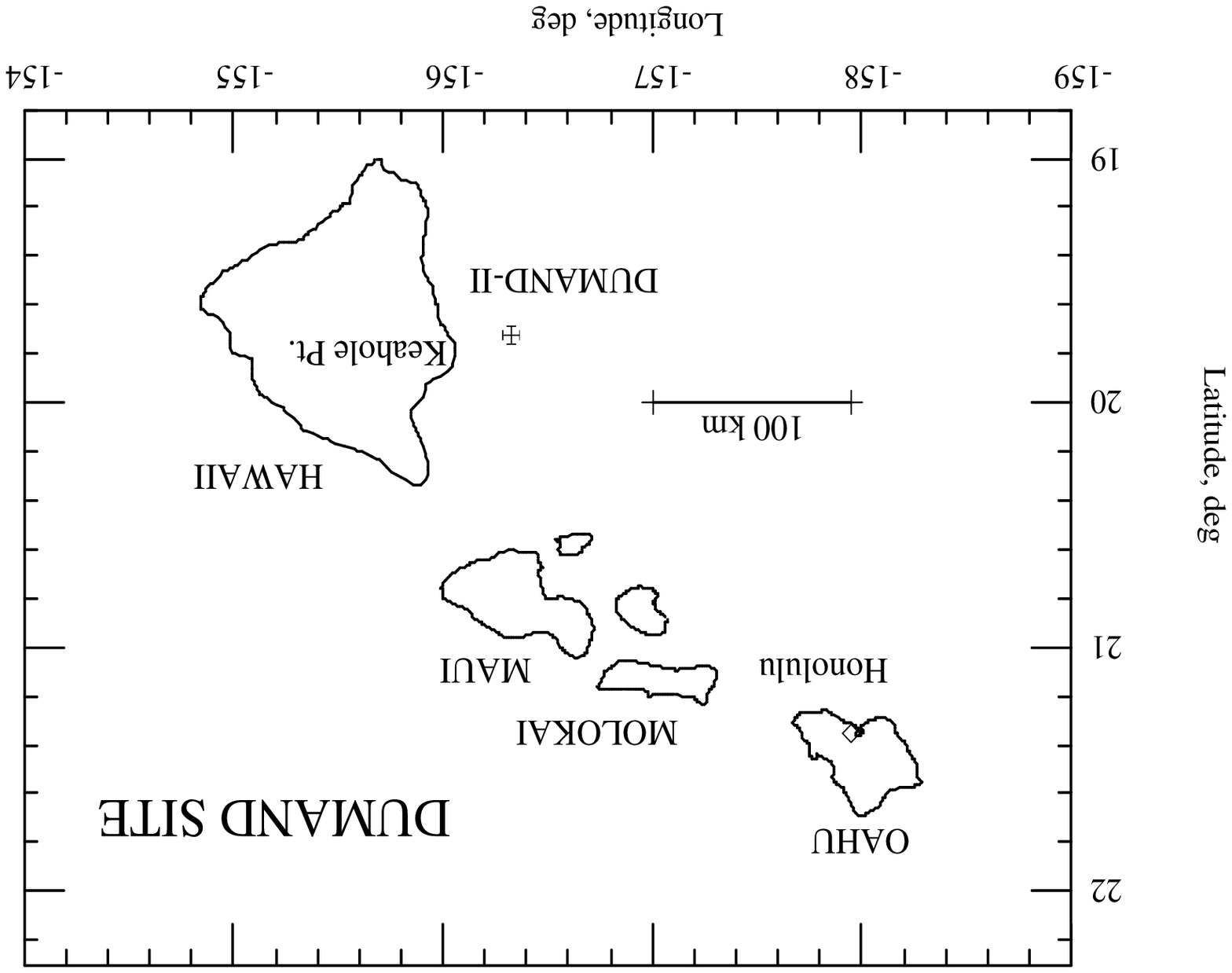}
\caption{DUMAND site off the Big Island of Hawaii.}
\label{sitemap}
\end{figure}

\begin{figure}
\epsfxsize=5.75truein
\epsffile{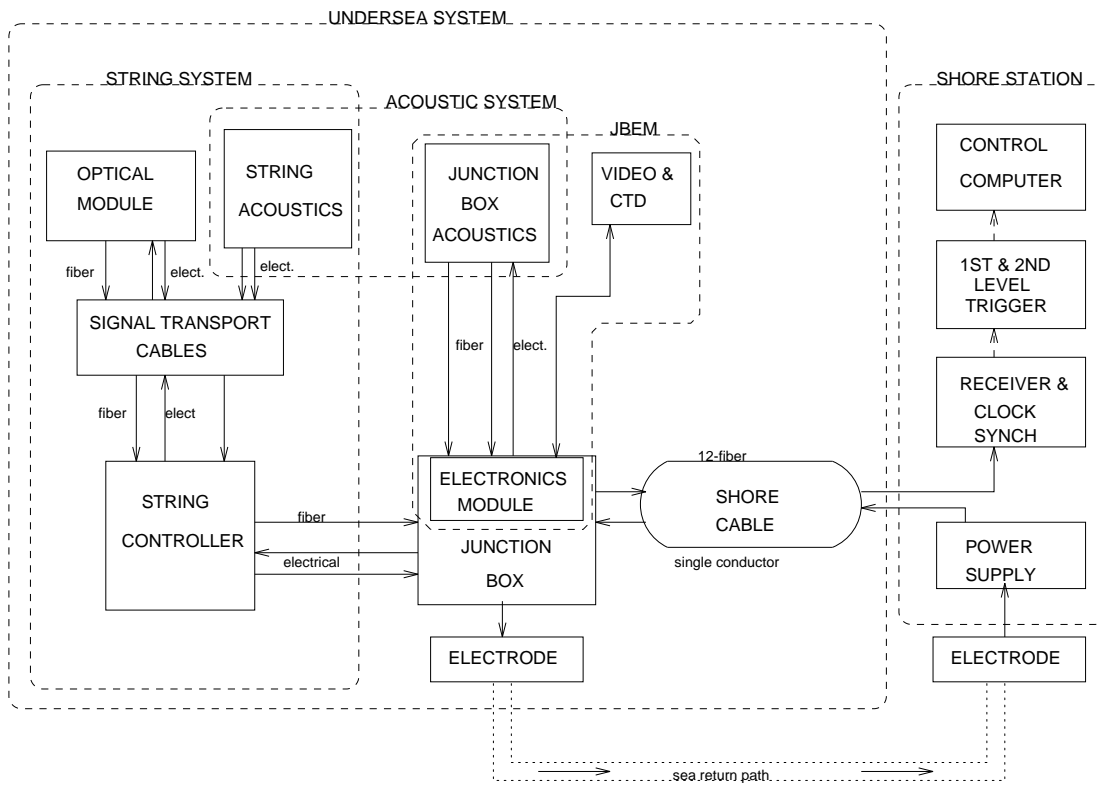}
\caption{Block diagram of the DUMAND detector system}
\label{dumblock}
\end{figure}

The array is being deployed on the ocean floor at depth 4800m, 30
km due west from the Kona Coast of the Island of Hawaii (Fig.~\ref{sitemap}),
and is connected to the shore laboratory at Keahole Point by a
cable combining electrical and fiber optic elements, terminating in
an underwater junction box. The shore cable contains 12 fibers
(including spares) and a copper layer which supplies 5 kW of
electrical power at 350 VDC, using a seawater return system.
Fig.~\ref{dumblock}
shows an overall block diagram for the DUMAND detector system. The
underwater site places no inherent limitation on possibilities for
future expansion of the detector.  With all 9 strings in place,
DUMAND will have an effective detection area of 20,000 m$^2$,
instrumenting a column of water which has the height of the Eiffel
tower and its width at the base.

\begin{figure}
\epsfxsize=5.75truein
\epsffile{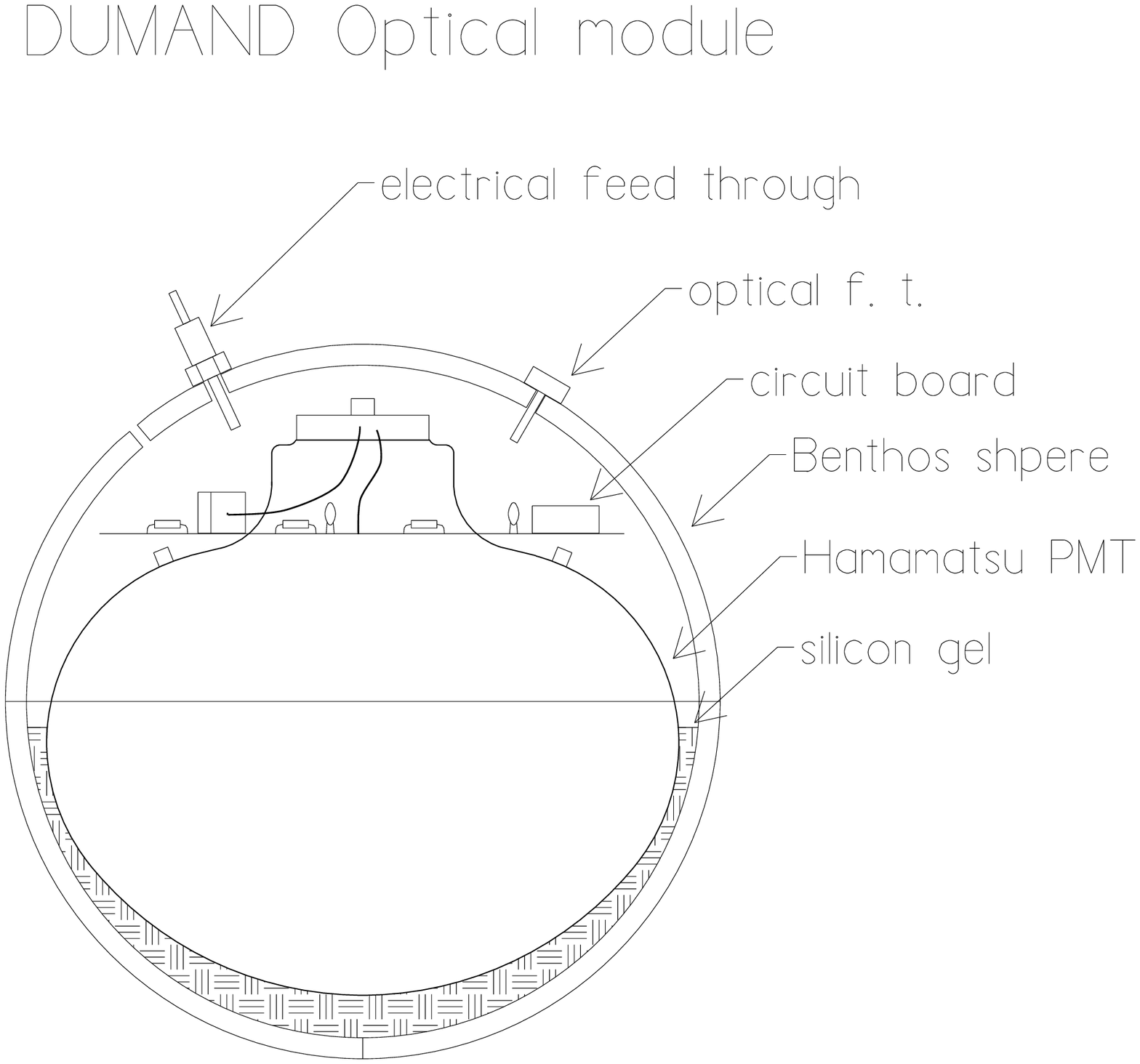}
\caption{DUMAND Optical Module.}
\label{omdraw}
\end{figure}

Signals from the PMTs are pre-processed within the optical modules
(Fig.~\ref{omdraw}),
providing standard pulses which encode time of arrival (to $\sim 1$ ns
accuracy), pulse area, and time-over-threshold (TOT), a measure of
pulse duration.  Data from  the 24 OMs on each string are digitized
and serialized in the string controller module by a custom 27
channel (including spares and housekeeping) monolithic GaAs
TDC/buffer/multiplexer chip which operates with 1.25 nsec timing
precision and 2-level internal buffers. The data stream is sent to
shore via optical fibers (one per string) at 0.5 GHz. A separate
optical fiber carries environmental and acoustical ranging
information which are used to measure the geometry of the array.

\begin{figure}
\epsfxsize=5.75truein
\epsffile{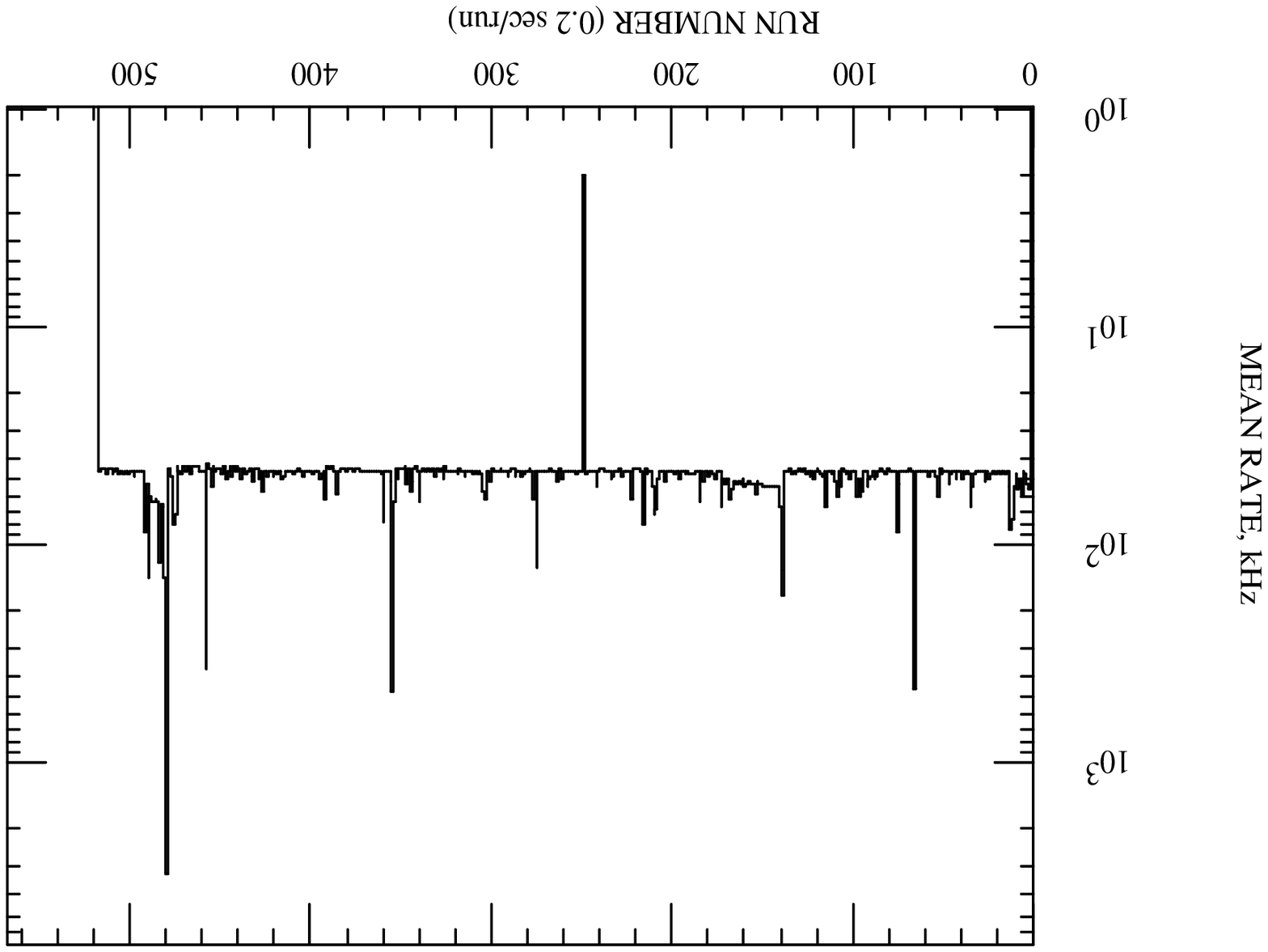}
\caption{Singles rates in a typical DUMAND optical module. The histogram shows
the mean counting rate over a series of 0.2 sec recording intervals. The
quiescent rate is about 60 kHz, with occasional intervals showing spikes above
100 kHz due to bioluminescence.}
\label{omrates}
\end{figure}

The data system has been designed to cope with the background rate
 from radioactivity in the water (primarily from natural $^{40}K$) and
bioluminescence and still generate minimal deadtime for
recording neutrino events.  Results from the 1993 deployment
confirmed observations made in the 1987 DUMAND-I
experiment\cite{bradner 1987}. As Fig.~\ref{omrates} shows, the dark counting
rate for a single
OM was found to be on the order of 60 kHz, primarily due to trace
$^{40}K$ in the huge volume of seawater each tube views. Noise due to
bioluminescence is episodic and likely to be unimportant after the
array has been stationary on the ocean bottom for some time, since
the light-emitting microscopic creatures are stimulated by motion.
$^{40}K$ and bioluminescence contribute mainly 1 photoelectron hits
distributed randomly in time over the entire array.

The raw information is sent to the shore station 30 km away for
processing.  The trigger system looks for patterns in time, space
and pulse height in the OM signals consistent with the passage of
charged particles through the array.  Events satisfying the trigger
criteria are recorded for further off-line analysis.

\begin{figure}
\epsfxsize=5.75truein
\epsffile{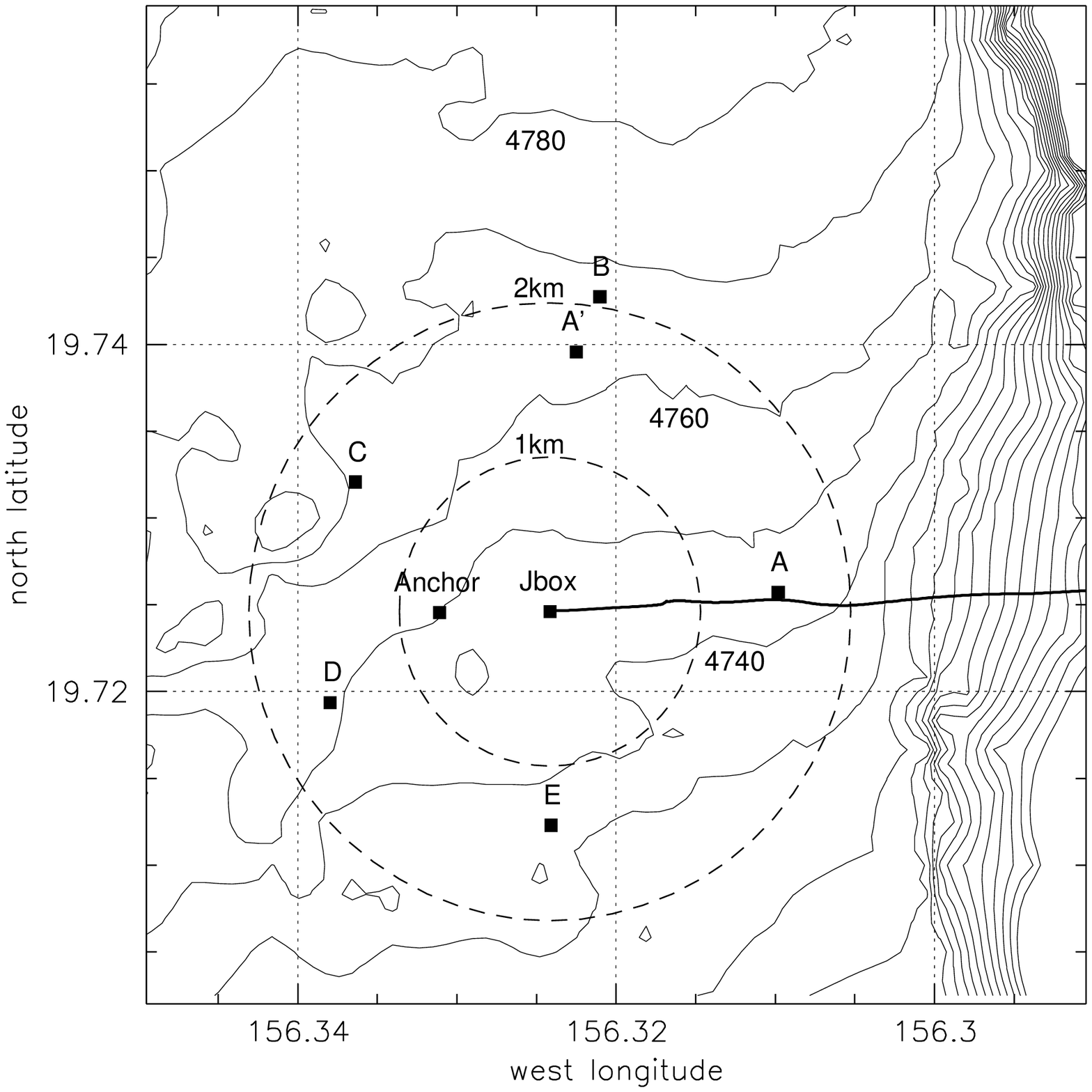}
\caption{Contour map of the DUMAND site (depths in meters below sea level),
showing placement of acoustical transponders, junction box, and cable as
surveyed during the 12/93 deployment operation.}
\label{site}
\end{figure}

Since  1992, DUMAND teams have been preparing the site and testing
underwater assembly  operations.  DUMAND-II  requires a reasonably
flat site with appropriate soil bearing properties.  The selected
site has been marked with acoustical transponders which have been
accurately surveyed in geophysical (GPS) coordinates (Fig.~\ref{site}), and its
suitability was verified remotely by acoustical imaging, film
camera and video recordings; in addition, DUMAND personnel have
cruised the area in a manned submarine, the US Navy's {\it DSV Sea
Cliff}, to  verify that the site is flat and free of any
undesirable features. These preliminary operations also confirmed
the exceptional clarity of the water, with attenuation length about
40m in the appropriate wavelength band.

We need to point reconstructed muon tracks onto the celestial
sphere with an accuracy better than 1$^o$ (the median angle between
primary $\nu$ and secondary $\mu$ at 1 TeV). This means that
relative OM locations must be known to the order of a few cm, and
the overall geographical orientation of the array must be known to
much better than 1 degree. The Global Positioning Satellite (GPS)
system plus conventional oceanographic acoustical survey techniques
allow us to measure the geographical coordinates of underwater
fiducials (acoustical transponders) to within a few meters,
satisfying the geographical orientation requirement.  We were
unable to find a commercial system able to reliably provide the OM
positioning accuracies required, so we designed our own sonar
system, which measures acoustical signal transit times with 10
$\mu$sec precision using frequency modulated chirps and matched
filtering via DSPs\cite{berns 1993}. Other components of the
environmental monitoring system measure oceanographic parameters
such as water currents, temperature and salinity (needed to
calculate the local speed of sound).

In December of 1993, the DUMAND scientific team and the crew of the
University of Washington oceanographic ship {\it R/V Thomas G.
Thompson} successfully deployed the first major components of
DUMAND, including the junction box, the environmental module, and
the shore cable, with one complete OM string attached to the
junction box.  Other DUMAND personnel prepared the shore station
for operation. The procedures for the lowering and cable laying
operations had been worked out in practice runs.  Cable laying
equipment was leased and mounted on the ship.  Environmental
monitoring equipment and the site-defining navigational sonar array
were also laid out and used in the deployment operation.

The basic infrastructure for DUMAND, comprising the underwater
junction box, the 30 km optical fiber/copper cable to shore, and
the shore station facility are now in place. The deployed string
was used to record backgrounds and muon events.  Unfortunately, an
undetected flaw in one of over 100 electrical penetrators
(connectors) used for the electronics pressure vessels produced a
small water leak.  Seawater eventually shorted out the string
controller electronics, disabling further observations after about
10 hours of operation. In January, 1994, the disabled string was
remotely released by an acoustical signal, recovered at sea, and
returned to Honolulu for diagnosis and repair.  The fault has been
analyzed and quality assurance procedures to avoid future recurrences have
been put in place.

In addition to the refurbished first string, two further strings
are currently  undergoing final assembly and testing.  We plan to
make extensive deep water tests of these three strings before
mooring them at the DUMAND site. Surface ship and underwater
vehicle resources needed to carry out deployment and
interconnection operations will be available in 1995.

After redeployment of the first string of OMs, each  successive
string will be moored at the vertices of an octagon at a radius of
40 m. Acceptable placement error is about 5m; this tolerance can be
readily achieved using available ships with dynamic positioning
capability (basically, GPS navigation coupled to the ship's
thrusters), according to simulation studies performed by a marine
operations consulting firm. Strings will be connected to the
junction box by an umbilical cable and wet-mateable
electrical/fiber-optic connector.  Using a mockup junction box and
string mooring, the US Navy's Advanced Tethered Vehicle (ATV)
carried out successful tests of the connecting operation in 1992,
proving that tethered remotely operated vehicles (ROVs, which are
cheaper and more readily available than manned submersibles) are
also an option for DUMAND underwater maintenance activities.

Although the success of the DUMAND deployment was marred by the
failure of a single penetrator, enough was learned from the
limited period of live operation to be confident that it will be
possible to complete and operate the whole DUMAND array. The
failure provided an undesired but nonetheless useful opportunity to
test procedures for recovering faulty equipment from the sea, an
essential task for long term operation.   The overall plan is to
install and operate three strings as a full-up demonstration, and
then proceed to deployment of the remaining six strings after about
a year of test operation.

Further information on DUMAND is available via the DUMAND Home Page
on the World Wide Web. The URL address is
\noindent
\begin{verbatim}
http://web.phys.washington.edu/dumand.html
\end{verbatim}

\section{AMANDA}

\begin{figure}[p]
\epsfxsize=5.75truein
\epsffile{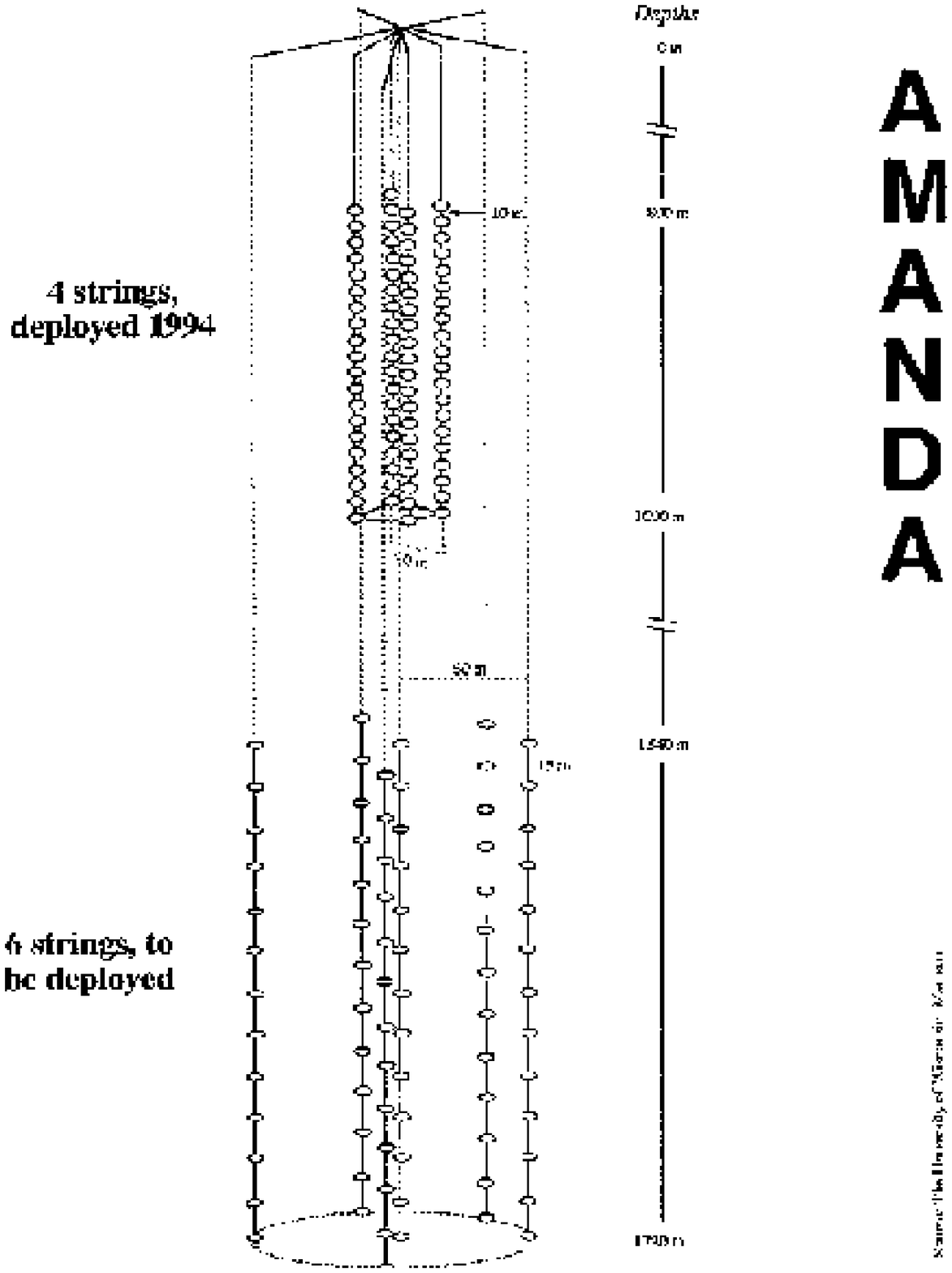}
\caption{AMANDA array: upper portion was deployed in 1/94, lower portion is to
be deployed in 12/95.}
\label{amanda}
\end{figure}

The Antarctic Muon and Neutrino Detector (AMANDA) uses the same
fundamental detector concept as DUMAND, but substitutes polar ice
for abyssal seawater\cite{amandacollab}. Photomultiplier tubes are
placed in vertical shafts melted into the icecap at the South Pole,
and data acquisition is handled in a counting house established at
the surface\cite{amandaicrc 1993}. The detector layout is depicted in
Fig.~\ref{amanda}.

This approach exploits two significant advantages of ice as a
medium: it is a stable solid, and it is biologically and
radiologically sterile. The ice forms a rigid, adaptive support for
the OM strings, and thus the need for measuring OM positions is
reduced from a continuous monitoring process to a one-time survey
procedure during deployment. Backgrounds due to bioluminescence and
natural radioactivity such as $^{40}K$ are effectively absent,
reducing the background noise rate substantially, and allowing
lower true event rates per sky pixel to be detected as a
significant excess\cite{amanda 1994}.

Only the Antarctic plateau provides a layer of ice of sufficient
depth, about 3 km total (although deployment depths are for
practical purposes limited to about 2 km). While real logistical
costs are very high, the US National Science Foundation operates a
vigorous, well-supported research program in Antarctica. One
significant result is ample support for the operational aspects of
AMANDA, from a source independent of conventional particle physics
funding. The
US South Pole Station is well equipped, and staffed year-round.
Access is by air only, and field operations can take place only
during the Austral summer season, roughly October through
February. A small staff of technicians and scientists volunteers to
remain icebound through the 6-month winter season, maintaining
experiments and forwarding limited amounts of data to the
continental USA via satellite links and land lines. While data
rates for communications will be improving over the next few years (plans exist
to provide the South Pole Station with 56 kB/sec Internet access)
AMANDA presently must depend to some extent on suitcases full of
tape cassettes for data transfer.

\begin{figure}[hp]
\epsfxsize=5.75truein
\epsffile{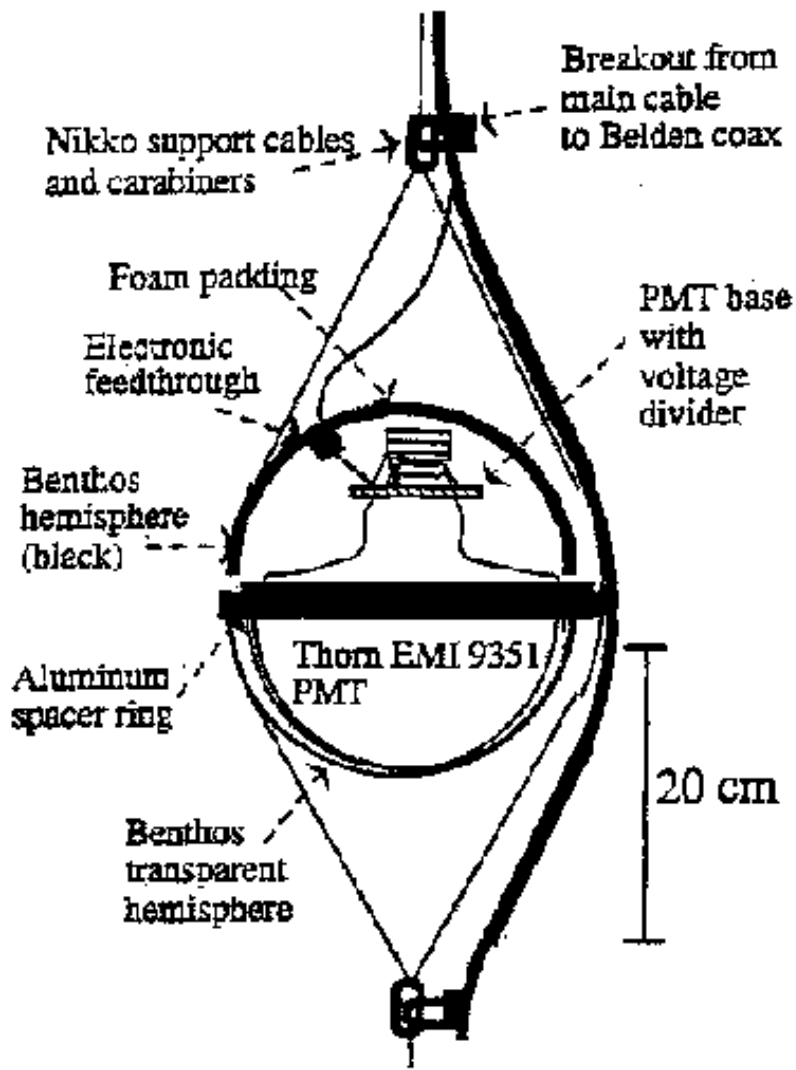}
\caption{AMANDA Optical Module.}
\label{amandaOM}
\end{figure}

Since the data acquisition system is only a short distance away from the
OMs, at the surface of the ice, AMANDA does not require front-end
electronics to be built into the optical modules or a local string
controller; the OMs, as shown in Fig.~\ref{amandaOM}, are thus just PMTs in a
glass pressure sphere
(the same type used in DUMAND), connected to the outside world by
coaxial cable (which also carries in the high voltage power
supply). Signal degradation produces some limitations on cable
length, but for the relatively shallow depths used thus far, and
planned for the next stage of deployment, there should be no
significant loss of timing information. The advantage of having
foolproof, simple, dumb OMs is very tangible.

The remote location, with highly limited access and long supply
lines, causes fewer difficulties than might be imagined, although
careful planning is essential (and enforced by Antarctic Program
management, who have long experience in these matters). One is
about 5,000 km from the nearest electronics parts store, and half
the useful season can be lost waiting for a forgotten item, so the
supply of spares and equipment must be thought through very
carefully and stringent predeployment testing is required.

An additional problem is the need for fuel to melt holes over one
km deep and about 60 cm in diameter for string deployment. The
initial deployments took advantage of a cache of surplus aircraft
fuel at the South Pole, stored too long to be certifiable for
aircraft use but perfectly suitable for ice-melting. This supply
has been consumed, and future deployments will require every liter
of fuel to be flown in (along with all other supplies). Since the
existing shafts (approximately 1 km deep) consumed about 12,000
liters of fuel each, and deeper shafts require disproportionately
larger amounts of fuel, this is a serious concern. However
experience from the initial operations led to a more efficient
drill design, now under construction, and it is expected that the
deeper holes now required can be made without substantially
increasing the fuel requirements.

A test string of four 20 cm diameter OMs was successfully deployed
and operated at 800 m depth in 1992. The PMTs used were available
 from a previous experiment, and OM size was limited by drilling
capabilities. Data on the flux of Cerenkov light from down-going
muons were interpreted to mean that the ice at $\sim 1$ km was
essentially bubble-free, and results from this test were considered
sufficiently promising to proceed to a first-stage deployment of
four full strings, each containing 20 OMs, in 1994. In this
operation, the drilling system performed very well, operating
nearly continuously for about 45 days and drilling holes at the
rate of 90 hr/km.

The OM signal characteristics from the 1994 deployment were about
as expected: timing resolution about 5 nsec, stable operation with
gain $10^8$, dark noise rate about 2 kHz. Of the 80 OMs deployed, 73
were operating well 5 months later, a reasonable survival rate. In
addition to coaxial cables carrying power down and signals up, the
strings included optical fibers to distribute calibration signals
 from a laser source on the surface to each OM. Each optical fiber
terminates in a nylon diffusing sphere located 30 cm from its OM.

\begin{figure}[hp]
\epsfxsize=5.75truein
\epsffile{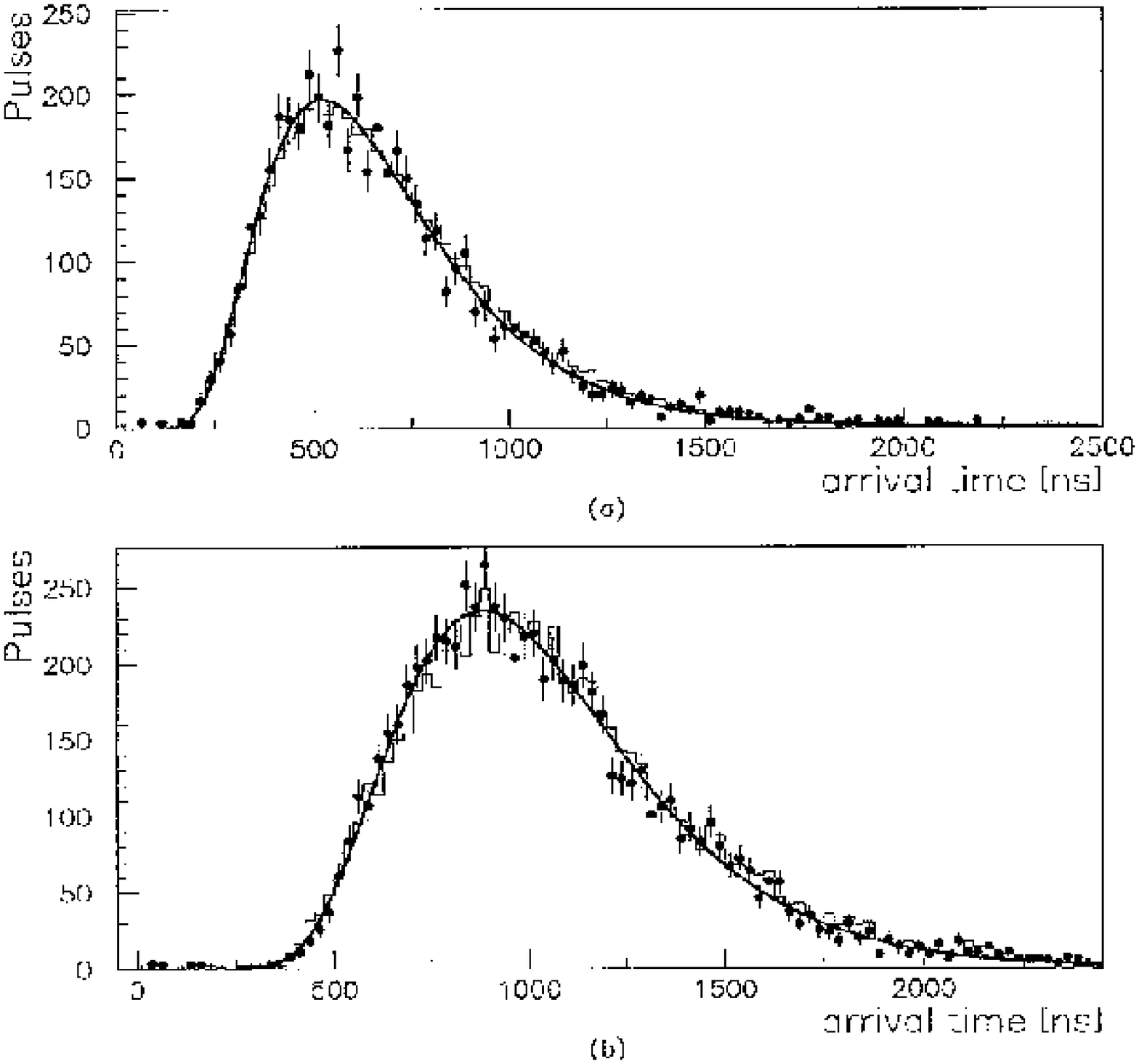}
\caption{Optical pulse transit time distributions from AMANDA calibration data,
for distances of a) 21 and b) 32 meters. The expected arrival times for direct
paths would be approximately 92 and 140 nsec respectively. Solid lines show
fits
to a diffusion model with appropriate effective scattering length (see
Fig.~\ref{amandascatt}).}
\label{amandatransit}
\end{figure}

\begin{figure}[hp]
\epsfxsize=5.75truein
\epsffile{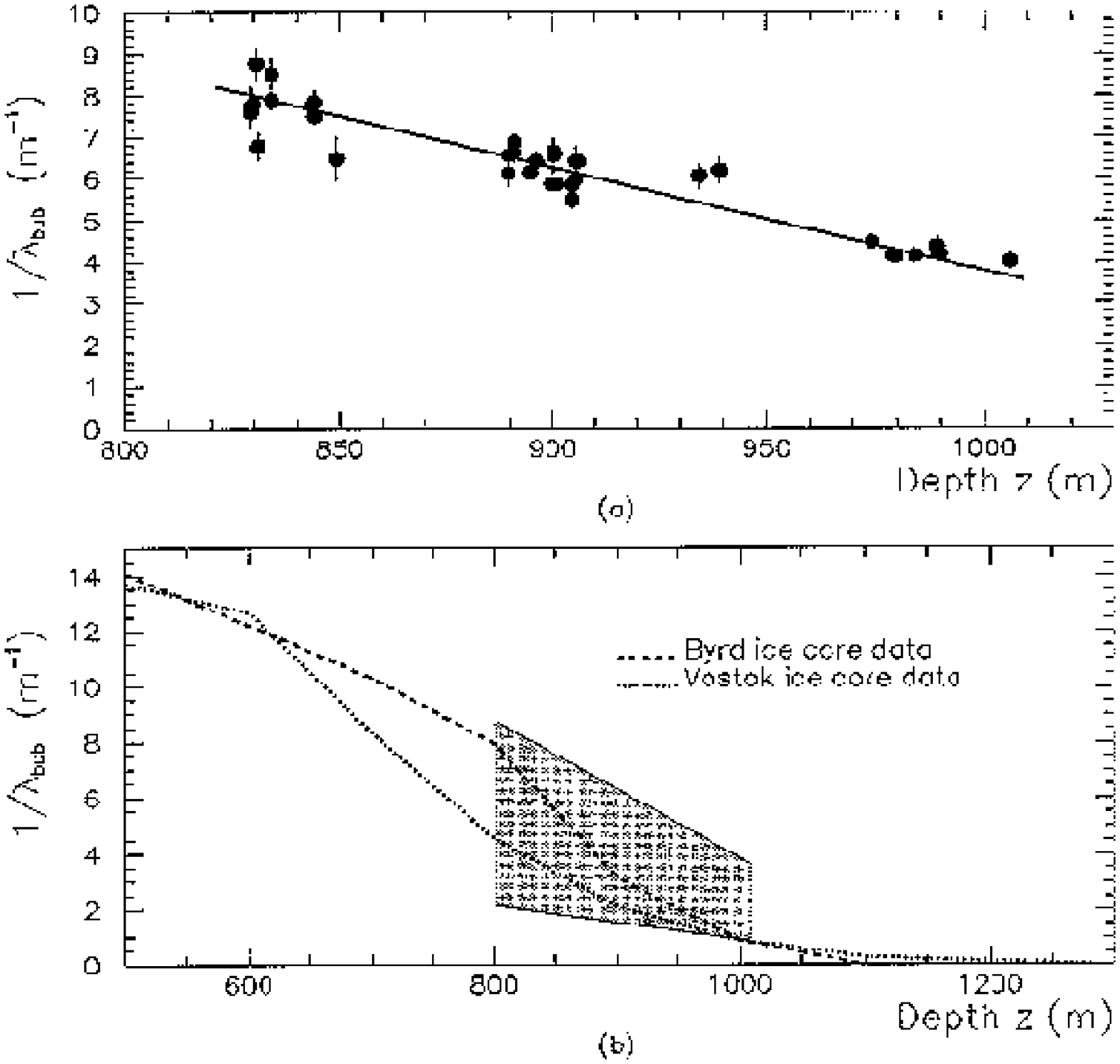}
\caption{Inverse effective scattering length for light at the AMANDA site as a
function of depth in ice.}
\label{amandascatt}
\end{figure}

Unfortunately, laser calibration signals were found to have transit
times between diffuser balls and OMs that were much longer than
expected for unobstructed straight-line paths. Fig.~\ref{amandatransit} shows
two
examples of transit time distributions, with the geometrical
distance between source and OM corresponding to arrival time delays
of 91 and 142 nsec respectively\cite{amanda 1994}.
As can be seen from the figure, the mean arrival time is more
than 5 times longer, and even
the earliest arrivals take nearly twice as long as expected to
reach the OMs. These data have been carefully analyzed by the
AMANDA group, and the conclusion is that a) the absorption length
of 475 nm photons in polar ice is about 60 meters, but b) the ice
contains a significant density of bubbles which produces an
effective scattering length of only 20 cm. Fig.~\ref{amandascatt} shows
that the arrival time data provide a good fit to these hypotheses.

The depth dependence of the scattering length is consistent with
results from microscopic examination of ice cores from Greenland
and Vostok (a Russian Antarctic base). At Vostok, where the
altitude and snow accumulation rate differ from the South pole, but
the ice temperature profile is similar, core samples show fewer
than 0.5 bubbles/cm$^3$ below 1280 meters depth. This gives hope that
putting the AMANDA strings only a few hundred meters deeper will
eliminate the scattering problem. The strategy will therefore be to
deploy the next set of strings in 1995-96, taking advantage of the
verified 60 m absorption length to increase OM spacing, and putting
the strings in below 1500 m to avoid bubbles. With an increase to
15 m vertical OM spacing,
a considerably larger volume can be instrumented. Six strings of 13
OMs each will be deployed in a circular pattern with 60 m radius.
The new drilling system may also make it possible to go to larger
diameter phototubes, although current plans call for using the same
PMTs used in previous deployments\cite{barwick 1994}.

As with DUMAND, the results of the 1994 AMANDA deployment did not
include detection of astrophysical neutrinos, but did demonstrate
important aspects of the technique. Despite the short scattering
length, which in effect reduces track reconstruction accuracy to
$\pm 10^o$ on the sky, it was possible to perform a number of tests
which verified the general viability of the AMANDA concept using
the 1994 array. AMANDA has much less overburden than DUMAND, and
therefore a much higher background rate due to downward-going
muons. However, the absence of bioluminescence and natural
radioactivity makes the OM singles noise rate much lower: about 2
kHz as compared to 60 kHz. The mean OM dark noise rates
observed (1.8 kHz) are about half what had been
anticipated.

Finally, it was possible to operate the strings in
coincidence with the South Pole Air Shower Experiment (SPASE),
which is located about 800 m away from the AMANDA site. Extensive
air showers arriving with zenith angles between 37 and 46 degrees
and with appropriate azimuth should be seen by both experiments,
and this mode of operation has been successfully demonstrated by
using SPASE triggers to log AMANDA data\cite{halzen 1994}.

Further information on AMANDA is available via the AMANDA Home Page
on the World Wide Web. The URL address is
\noindent
\begin{verbatim}
http://spice2.physics.wisc.edu/amanda2.html
\end{verbatim}

\section{Comparison of AMANDA and DUMAND}

The following table compares salient features of the two detectors.
In addition to common features, both projects have a set of unique
advantages and disadvantages, often in the form of a tradeoff. For
example, AMANDA has rigidly fixed OM positions and the ability to
locate front-end electronics very near the detector elements, on
the surface just above the array. On the other hand, DUMAND strings
can be readily released and recovered for repair or repositioning,
and the use of fiber optic data transmission makes cable length
irrelevant. DUMAND's thick seawater overburden greatly reduces
event backgrounds due to down-going muons, at the expense of
heavier singles rates due to radioactivity and bioluminescence,
while AMANDA's ice overburden is less than half as thick but makes
no contribution to dark noise. The real costs of deployment are
probably about equal, but AMANDA's logistical costs are part of a
very large Antarctic research enterprise in which AMANDA is (at
present) a small perturbation, while DUMAND's costs are a very
visible portion of their budget (although in fact ship and
submarine time should eventually be available by interagency
cooperation). The two groups have had similar outcomes from their
first major deployment attempts this year: partial proof of
concept, but not the definitive proof offered by unambiguous
neutrino detection.

\begin{table}[hp]
\caption{\bf COMPARISON BETWEEN DUMAND AND AMANDA}
\label{dumanda}
\medskip
\begin{tabular}{|ll|}
\hline
\bf DUMAND & \bf AMANDA\\
&\\
\bf Seawater -- high noise & \bf Ice -- low noise\\
\hspace{.25in}$\bullet$ $^{40}K$ background &
\hspace{.25in}$\bullet$ No $^{40}K$ background\\
\hspace{.25in}$\bullet$ Bioluminescence & \hspace{.25in}$\bullet$
No bioluminescence\\
 & \\
\bf Deep: 5000 m & \bf Shallow: 1000 m\\
\hspace{.25in}$\bullet$ Low event background &
\hspace{.25in}$\bullet$ High event background\\
\hspace{.25in}$\bullet$ Smart OMs & \hspace{.25in}$\bullet$
Simple OMs\\
\hspace{.25in}$\bullet$ Digital fiber-optic data transfer &
\hspace{.25in}$\bullet$ Analog signals to surface -
coax cable\\
\hspace{.25in}$\bullet$ Complex underwater electronics  &
\hspace{.25in}$\bullet$ Simple OMs - processing on surface\\
 & \\
\bf Underwater  & \bf Under ice\\
\hspace{.25in}$\bullet$ Track visibility proven &
\hspace{.25in}$\bullet$ Bubbles remain at 1000m\\
\hspace{.25in}$\bullet$ Well-developed commercial&
\hspace{.25in}$\bullet$ Environment less well known\\
\hspace{.35in} technologies & \\
\hspace{.25in}$\bullet$ DSV/ROV required &
\hspace{.25in}$\bullet$ Direct access from surface\\
\hspace{.25in}$\bullet$ Recoverable after deployment &
\hspace{.25in}$\bullet$ Not recoverable once deployed\\
 & \\
\bf Hawaii & \bf Antarctica \\
\hspace{.25in}$\bullet$ Easy access year-round &
\hspace{.25in}$\bullet$ Restricted access to site\\
\hspace{.25in}$\bullet$ Local high-tech facilities &
\hspace{.25in}$\bullet$ Limited facilities at site\\
\hspace{.25in}$\bullet$ Local university group &
\hspace{.25in}$\bullet$ No permanent residents \\
\hspace{.35in}(resident staff planned) &
\hspace{.35in}(but continuous staffing) \\
\hspace{.25in}$\bullet$ Near-equatorial site: daily &
\hspace{.25in}$\bullet$ Polar site: fixed view of \\
\hspace{.35in} scan of celestial mid-latitudes &
\hspace{.35in} celestial northern hemisphere\\
 & \\
\hline
\multicolumn{2}{c}{\bf Common Features:}\\
\multicolumn{2}{c}{Same basic techniques used}\\
\multicolumn{2}{c}{Overall costs $\sim$ same}\\
\multicolumn{2}{c}{Site permits expansion to next-generation
size (1 km$^3$)}\\
\end{tabular}

\end{table}

While both DUMAND and AMANDA are pursuing the Cerenkov light
technique, earlier investigations have suggested that a very large
volume detector of high energy neutrinos can be constructed at very
low cost using acoustical detection.  The deposition of energy in
the water by produced particles generates a low level
characteristic bipolar sound pulse with an effective frequency
spectrum peaked in the range 30 to 60 KHz.  The hydrophone array
built into DUMAND for its positioning system is very efficient in
this range, and should be capable of detecting particle cascades of
about 1 PeV at a range of 40m\cite{learned 1993}. Simulation studies
suggest that by using noise cancellation and signal coherence
techniques (ie, treating our set of hydrophones as a phased array),
it will be possible to systematically enhance noise rejection and
detect high energy particles.  The DUMAND array will be equipped
to observe coincidences of OM and acoustical signals and this will
provide the first direct practical test of acoustical detection.
DUMAND will also supply acoustical equipment to AMANDA for tests of
acoustical detection in the ice.

Throughout the process of detector construction and deployment, the
two groups have engaged in mutual assistance and cooperation
despite the inevitable sense of competition.  It is quite likely
that at some point in the future we will be working together
directly to focus resources and expand capabilities. The present
DUMAND and AMANDA arrays, even after all currently planned
deployments are completed, will serve primarily as test beds and
prototypes for a much larger detector.

\section{The Next Step: km$^3$}

Both the DUMAND and AMANDA groups acknowledge that detectors with
effective areas on the order of $10^4$ m$^2$ provide marginal
capability for detecting neutrino sources given present theoretical
estimates as well as data on gamma rays. The aim of the present
generation of detectors, including Baikal and Nestor, is to
demonstrate the value of neutrino astronomy by providing the first
look at the neutrino sky. Definitive results will be likely to come
 from the next generation of neutrino detectors, which must have
sensitive volumes on the order of a cubic kilometer. Given the history of
DUMAND, with a delay of nearly 30 years between the first
discussions of the detector concept and its materialization in
hardware, everyone with an interest in neutrino astronomy is
concerned about reducing the lead time for the next step. In part,
during the early years DUMAND was a concept waiting for the
development of appropriate technology ({\it eg}, wet-mateable fiber
optics connectors, which became available in the late 80s), but it
is certainly not too early to begin design and organizational
activities on the second generation now.

It seems clear that both the deep-sea and polar-ice approaches have
valuable features as well as problems that are not yet resolved, at
least to the satisfaction of the community at large. At present it
is still possible that AMANDA will find no end to its bubble
problem at practical depths. Similarly, although the basic
feasibility and technological issues are resolved, it is essential
for DUMAND to definitively demonstrate its ability to overcome
component reliability problems and operate a complex detector
system deep underwater on a long-term basis. If either group fails
to achieve these goals, the direction for future work will be
clear; in the happy circumstance that both detectors work as
planned, a decision about whether the km$^3$ detector should be
underwater or in the ice will be based on assessment of results
 from initial runs.

Several significant initiatives took place in early 1994: a
workshop held at the Jet Propulsion Laboratory led to the formation
of a US-based coalition to pursue a cubic kilometer detector, and
later the European Community's Megascience Workshop resulted in a
similar European coalition. At the 1994 Snowmass Summer Study
(entitled ``Particle and Nuclear Astrophysics and Cosmology in the
Next Millenium") an interest group combining both coalitions was
organized. The existing BAND groups (Baikal, AMANDA, Nestor,
DUMAND) are working with the JPL group and others to organize
workshops aimed at preparing a conceptual proposal before the end
of 1995, so that funding initiatives can begin promptly. Already,
JPL workers have begun development of new OM designs which have
extremely low power consumption and use optical fibers for power as
well as data transmission. Interested individuals should join the
group to keep apprised of progress; consult the Worldwide Web
for further details:
\begin{verbatim} http://web.phys.washington.edu/km3.html
\end{verbatim}

\section{Acknowledgments}

Special thanks are due to Steve Barwick, Francis Halzen, John
Learned, and Robert Morse for help in assembling the information needed for
this review. However, blame for mistakes and invalid opinions expressed
lies with me alone. Thanks are also due to the SLAC Summer School
organizers and staff.

\end{document}